\documentclass[aps,prb,twocolumn,superscriptaddress]{revtex4}
\usepackage[latin9]{inputenc}
\usepackage{graphicx}
\usepackage{amsmath}
\usepackage{float}
\usepackage{hyperref}

\begin{document}

\title{Minimally entangled typical thermal states with auxiliary matrix-product-state bases}

\author{Chia-Min Chung}
\affiliation{Department of Physics and Arnold Sommerfeld Center for Theoretical Physics,
Ludwig-Maximilians-Universit\"{a}t M\"{u}nchen, 80333 Munich, Germany}
\affiliation{Munich Center for Quantum Science and Technology (MCQST), Schellingstrasse 4, 80799 Munich, Germany}
\author{Ulrich Schollw\"{o}ck}
\affiliation{Department of Physics and Arnold Sommerfeld Center for Theoretical Physics,
Ludwig-Maximilians-Universit\"{a}t M\"{u}nchen, 80333 Munich, Germany}
\affiliation{Munich Center for Quantum Science and Technology (MCQST), Schellingstrasse 4, 80799 Munich, Germany}

\begin{abstract}
Finite temperature problems in the strong correlated systems are important but challenging tasks.
Minimally entangled typical thermal states (METTS) are a powerful method in the framework of tensor network methods to simulate finite temperature systems, including Fermions and frustrated spins which introduce a sign problem in the typical Monte Carlo methods.
In this work, we introduce an extension of the METTS algorithm by using a new basis, the auxiliary matrix product state.
This new basis achieves the pre-summation process in the partition function, and thus improve the convergence in the Monte Carlo samplings.
The method also has the advantage of simulating the grand canonical ensemble in a computationally efficient way by employing good quantum numbers.
We benchmark our method on the spin-$1/2$ XXZ model on the triangular lattice, and show that the new method outperforms the original METTS as well as the purification methods at sufficiently low temperature, the usual range of applications of METTS.
The new method also naturally connects the METTS method to the purification method.
\end{abstract}

\maketitle

\section{Introduction}
Finite-temperature problems in the strongly correlated systems are important, but challenging tasks in condensed matter physics.
Many interesting systems can be investigated only via numerical simulations.
Quantum Monte Carlo is one of the most common methods, but it typically encounters the minus-sign problem for Fermionic and frustrated-spin systems, and thus is limited in the applications at low temperature.
Inspired by the density matrix renormalization group (DMRG)~\cite{PhysRevLett.69.2863,PhysRevB.48.10345,RevModPhys.77.259},
matrix product states (MPS)~\cite{McCulloch_2007,Schollw_ck_2011} have been introduced to represent low entanglement states, and have been shown to be excellent approximations of the ground states of local Hamiltonians in low dimensions.
Based on MPS, several finite-temperature methods have been developed, including minimally entangled typical thermal states (METTS)~\cite{PhysRevLett.102.190601,Stoudenmire_2010,PhysRevB.92.115105}, the purification method~\cite{PhysRevLett.93.207204,PhysRevLett.93.207205,PhysRevB.72.220401,PhysRevB.93.045137,PhysRevB.94.115157,PhysRevB.98.235163}, and exponential tensor renormalization group (XTRG)~\cite{PhysRevX.8.031082,PhysRevB.100.045110} (and the method based on transfer matrix renormalization group (TMRG)~\cite{doi:10.1143/JPSJ.66.2221,doi:10.1143/JPSJ.64.3598,Bursill_1996,PhysRevB.56.5061} while we will not discuss in detail). 

These methods build on representing the density operator as a matrix product operator or its analogues, and the approximation relies on that the density operator has low entanglement.
Purification and XTRG anneal the full density operator from infinite temperature to the target temperature and obtain the finite temperature density operator.
The METTS method stochastically samples the partition function by an ensemble of states,
and thus involves the annealing only of pure states rather the density operator.
Since at high temperature the entanglement of the density operator is small, the purification and XTRG are expected to be more efficient~\cite{PhysRevB.92.125119} because they don't need stochastic sampling and have no statistical noise.
However at low temperature, where the entanglement is assumably large, METTS is expected to be more efficient, because it deals only with pure states which have significantly lower entanglement than the full density operator.
At zero temperature, the density operator is an outer product of the ground state, and thus has double entanglement than the ground state, squaring the required bond dimension.

In this work we introduce an extension of the METTS method by introducing a new type of basis, the auxiliary MPS (AMPS).
The new basis achieves the pre summation process in the decomposition of the partition function, and thus improves the convergence in the Monte Carlo samplings.
The method also has the advantage of simulating the grand canonical ensemble using quantum-number (QN) conservation, which is important to reduce the computational cost.
The use of QN has been achieved in Ref.~\cite{PhysRevB.95.195148}, however in a rather complicated way.
We demonstrate that our method is more efficient than both the original METTS and the purification method at low temperatures by comparing the convergence properties of the energy and the correlations.
Our method also provides a nice connection between the METTS and the purification methods.

\begin{figure}[h!]
\includegraphics[width=1\columnwidth]{{{p}}}
\caption
{
    (a) The probability of collapsing a MPS from $|\phi_i\rangle$ to $|i'\rangle$.
    Since $|\sigma\rangle\langle\sigma|=\hat{1}$, the most of terms cancel and the final probability $\propto |\langle i'|\phi_i\rangle|^2$.
    (b) The probability of collapsing a AMPS. Similarly the final probability $\propto |\langle i'|\phi_i\rangle|^2$.
}
\label{fig:prob}
\end{figure}

\begin{figure}[h!]
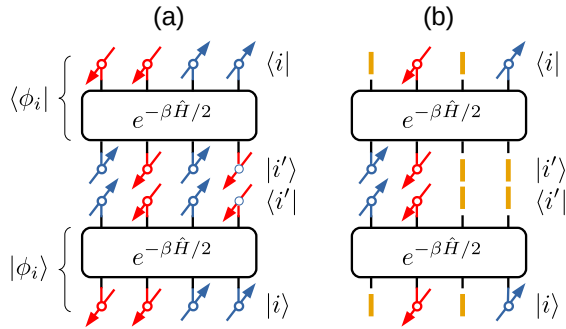

\includegraphics[width=0.9\columnwidth]{{{conf}}}
\caption
{
    (a) An example of the configurations in the METTS algorithm.
    $|i\rangle$ and $|i'\rangle$ are product states and can be represented by MPS with bond dimensions $1$.
    (b) An example of the configurations in the new algorithm with AMPS bases.
    $|i\rangle$ and $|i'\rangle$ are product AMPS of bond dimensions $1$ with two auxiliary indices.
    The tensors on the uncollapsed sites are updated as the identity operators (orange lines).
}
\label{fig:conf}
\end{figure}

\section{METTS algorithm in configuration representation}
\label{sec:METTS}
In this section we present the original METTS algorithm in a way different from the original way, but equivalent to it.
More precisely we represent the algorithm by sampling configurations, or "diagrams".
The new representation is useful to generalize the algorithm and develop new methods.

\subsection{Original representation}
\label{sec:orig}
The original METTS algorithm is described as follows. Starting from a random product state $|i\rangle$, one repeats the following two steps:
1) Compute $|\phi_i\rangle=e^{-\beta \hat{H}/2}|i\rangle / C$ where $C$ is the normalization constant, and 
2) collapse $|\phi_i\rangle$ to a new product state $|i'\rangle$ with probability $p(i\to i')=|\langle i'|\phi_i\rangle|^2$.
The collapsing step will be explained below.
By repeating these two steps, one obtains $|i\rangle$ with probability $\propto \langle i|e^{-\beta \hat{H}}|i\rangle$, which samples the Boltzmann distribution.
The measurements are taken on $|\phi_i\rangle$ in each sampling, and the expectation values are approximated by the Monte Carlo mean values.

The imaginary time evolution in step 1) can be performed, for instance, by using time-evolving block decimation (TEBD)~\cite{PhysRevLett.93.040502,PhysRevLett.93.207204,PhysRevLett.93.076401,Daley_2004} or the time dependent variational principle (TDVP)~\cite{PhysRevLett.107.070601,PhysRevB.94.165116}.
The detail of the time evolution computation is not the focus of this work~\cite{Stoudenmire_2010,2019arXiv190105824P}.
In this work we use TEBD for all the simulations.

The collapsing step is done by collapsing site by site.
Without loss of generality, we consider $|i\rangle=|\sigma_1\rangle|\sigma_2\rangle\cdots|\sigma_N\rangle$ as a product state in the $S_z$ eigenbasis, where $|\sigma_k\rangle=\{\uparrow,\downarrow\}$.
To collapse the first site, one first computes the reduced density matrix $\rho_1$ for the first site.
The first site is then collapsed to a specific spin direction $|\bar{\sigma}_1\rangle$ with probability $\langle \bar{\sigma}_1|\rho_1|\bar{\sigma}_1\rangle/\sum_{\sigma_1=\uparrow,\downarrow}\langle \sigma_1|\rho_1|\sigma_1\rangle$.
After obtaining the first state $|\bar{\sigma}_1\rangle$, one projects the first site of $|\phi_i\rangle$ to $|\bar{\sigma}_1\rangle$; the resulting state is denoted by $|\phi_i^{\bar{\sigma}_1}\rangle\equiv|\bar{\sigma_1}\rangle\langle\bar{\sigma}_1|\phi_i\rangle$.
Following the same procedure, the second site is collapsed to $|\bar{\sigma}_2\rangle$ with probability $\propto \langle \bar{\sigma}_2|\rho_2^{\bar{\sigma}_1}|\bar{\sigma}_2\rangle$, where $\rho_2^{\bar{\sigma}_1}$ is the second-site reduced density matrix of $|\phi_i^{\bar{\sigma}_1}\rangle$.
The collapsing steps are performed site by site until all the sites are collapsed, which generate a new product state $|i'\rangle$.
This procedure samples $|i'\rangle$ from $|\phi_i\rangle$ with the desired probability $|\langle i'|\phi_i\rangle|^2$, which can be seem in Fig.~\ref{fig:prob}(a).
The reader may want to find more detail of the original representation in Ref.~\cite{PhysRevLett.102.190601,Stoudenmire_2010}.

\subsection{Configuration representation}
Here we represent the METTS algorithm in a different picture.
Similar to general quantum Monte Carlo methods, we decompose the partition function by inserting two complete sets at $\tau=0$ and $\beta/2$
\begin{equation}
    Z = \sum_{|i\rangle|i'\rangle} W(|i\rangle,|i'\rangle),\quad W(|i\rangle,|i'\rangle) \equiv \langle i| e^{-\beta \hat{H}/2} |i'\rangle \langle i'| e^{-\beta \hat{H}/2} |i\rangle,    
\end{equation}
where $\sum_{|i\rangle|i'\rangle}$ sums over the complete-sets states $\{|i\rangle\}$ and $\{|i'\rangle\}$.
Without loss of generality, we again consider $|i\rangle$ and $|i'\rangle$ as product states in $S_z$ eigenbasis.
We represent each $W(|i\rangle,|i'\rangle)$ as a configuration, or a "diagram", as shown in Fig.~\ref{fig:conf}(a).
Each configuration is uniquely defined by $2N$ spins, where $N$ is the number of sites.
We then represent the METTS algorithm in sampling such configurations with probability $\propto W(|i\rangle,|i'\rangle)$.

The sampling process is done by sampling $|i\rangle$ and $|i'\rangle$ iteratively.
Without loss of generality, we first sample a new $|i'\rangle$ with fixed $|i\rangle$.
The sampling process is exactly the same with which described in Sec.~\ref{sec:orig}:
First compute $|\phi_i\rangle=e^{-\beta \hat{H}/2}|i\rangle / C$ where $C$ is the normalization constant, and then collapse to a new product state $|i'\rangle$.
The probability of having $|i'\rangle$ is $\propto |\langle i'|\phi_i\rangle|^2$, which is exactly the weight of the configuration $W(|i\rangle,|i'\rangle)$.
We then fix $|i'\rangle$ and sample a new $|i\rangle$, and so on.
The process is repeated until the enough configurations are sampled.

To compute the expectation value $\mathrm{Tr}(\hat{O} e^{-\beta\hat{H}})/Z$ of an observable $\hat{O}$, one can in principle measure at any imaginary time because of the trace.
In the METTS algorithm we measure at $\tau=0$ and $\tau=\beta/2$, which means measuring $O_i=\langle\phi_i|\hat{O}|\phi_i\rangle$ and $O_{i'}=\langle\phi_i'|\hat{O}|\phi_i'\rangle$.
The expectation value is then approximately obtained by the Monte Carlo mean value of $(O_i+O_{i'})/2$.

One can see that the algorithm described above is exactly the same with the original METTS algorithm, but with a new representation of sampling configurations.
This new representation is useful to generalize the algorithm and develop new algorithms.

\begin{figure}[h!]
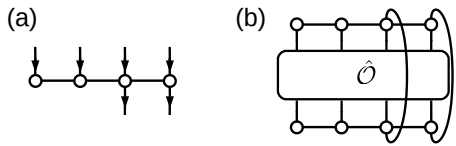

\includegraphics[width=0.7\columnwidth]{{{amps}}}
\caption
{
    (a) An AMPS with auxiliary indices on the two rightmost sites.
    (b) Measure an observable $\langle\phi_i|\hat{O}|\phi_i\rangle$ on a AMPS $|\phi_i\rangle$.
}
\label{fig:amps}
\end{figure}

\subsection{QN and bases choices in METTS}
The choice of bases $\{|i\rangle\}$ and $\{|i'\rangle\}$ is completely flexible, and $\{|i\rangle\}$ and $\{|i'\rangle\}$ are not necessarily the same.
A common choice is to use $|i\rangle$ in $S_z$ eigenbasis and $|i'\rangle$ in the $S_x$ eigenbasis, which we call it $S_z$-$S_x$ bases.
Such a choice can greatly reduce the autocorrelation and improve the Monte Carlo convergence~\cite{Stoudenmire_2010}.
For the pure $S_z$ basis, the sampling will be very inefficient at high temperature and/or with weak off-diagonal coupling.
It would be completely stuck at infinite temperature or in the absence of off-diagonal coupling.

Another important effect of the $S_z$-$S_x$ basis is on the QN.
If the Hamiltonian conserves, for example, total $S_z$ QN, and if we use pure $S_z$ basis that also conserves the total $S_z$ QN, then the whole simulation will stay in the same QN sector of the initial $|i\rangle$ and $|i'\rangle$, resulting in a canonical (ensemble) simulation.
This is sometimes desirable, as one wants to focus on a particular QN sector at low temperature.
However it is sometimes important to be able to fluctuate between all the QN sectors, especially at finite temperature where different QN sectors can contribute significantly.
By using the $S_z$-$S_x$ bases, one can simulate in the grand canonical ensemble because the $S_x$ basis provides the fluctuation between different total $S_z$ sectors.

The drawback of the $S_z$-$S_x$ bases is that, in general one can no longer use a QN conserving MPS.
(An exception occurs when the Hamiltonian is $SU(2)$ symmetric, for which one can rotate between $S_z$ and $S_x$ bases without changing the form of the Hamiltonian.)
The QN in a MPS is crucially important to reduce the computational cost by the block-diagonal structure it induces.
Therefore we would want to have an algorithm that simulates in the grand canonical ensemble but also conserves the QNs, which leads to our new algorithm.

\begin{figure}
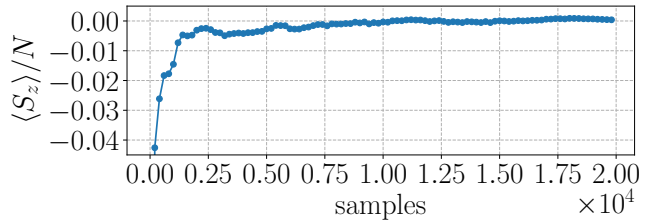

\includegraphics[width=1\columnwidth]{{{sz_L64_beta2}}}
\caption
{
    Convergence of the total $\langle S_z\rangle/N$ by using $N_\mathrm{aux}=2$.
    The system is a $L=64$ Heisenberg chain at the inversed temperature $\beta=2$.
    The initial state is the fully polarized state.
}
\label{fig:sz}
\end{figure} 

\section{METTS with auxiliary-MPS basis\label{sec:algorithm}}

\subsection{Algorithm}
In this section we introduce a new method to perform grand canonical simulations with QN conserved MPS.
To the authors' knowledge, Ref.~\cite{PhysRevB.95.195148} is the only work so far addressing this problem.
Here we provide a conceptually and implementationally simpler, and presumably more efficient way to achieve the same purpose.
The new method basically follows the original algorithm.
The key modification is in the collapsing step:
One does not collapse all the sites.
Instead one collapses only $N-N_\mathrm{aux}$ sites, and remains the other $N_\mathrm{aux}$ sites as \emph{local identity} operators.
The collapsing procedure generates a MPS with $N_\mathrm{aux}$ additional auxiliary site indices, as shown in Fig.~\ref{fig:amps}(a).
We call this kind of MPS an auxiliary MPS (AMPS), which can be also understood as partially projected matrix product density operators.
The configurations $W(|i\rangle,|i'\rangle)$ with AMPS $|i\rangle$ and $|i'\rangle$, and their corresponding weights, can be defined in the same way as before, as shown in Fig.~\ref{fig:conf}(b).

The positions of the uncollapsed sites can be fixed in the whole simulation or can be randomly chosen in each sampling.
If we consider fixed uncollapsed positions,
the sum of $W(|i\rangle,|i'\rangle)$ clearly still represents a decomposition of the partition function.
Therefore the algorithm remains the same as described in Sec.~\ref{sec:METTS}, in the way of sampling the configurations with probability proportional to the corresponding weights.
The technical detail of working with AMPS instead of MPS will be discussed.
Since every choice of the fixed positions results in the same partition function, one can average over all the possible choices.
As a result, one can choose the uncollapsed sites randomly in each sampling.
In this work we always choose the uncollasped sites randomly to reduce the autocorrelation.

An AMPS works almost the same as an MPS in the algorithm.
Here we summarize three places that involve the operations on AMPS.
1) In TEBD, the time evolution gates apply only on the physical indices, and thus the auxiliary indices remain uncontracted.
2) In the collapsing step, the reduced density matrices are computed by contracting both the physical and auxiliary indices.
As shown in Fig.~\ref{fig:prob}(b), the AMPS collapsing still provides the desired probability $p(i\to i') \propto |\langle i'|\phi_i\rangle|^2$ proportional to the configuration weights.
3) In measurements, both the physical and auxiliary indices are contracted, as shown in Fig.~\ref{fig:amps}(b).

\subsection{Quantum number and the convergence}
The auxiliary indices in the AMPS naturally provide fluctuations between the different QN sectors.
At the same time the whole AMPS is still QN conserved, so one can work with QN conserved AMPS.
The QN flow in the AMPS is indicated by the arrows in Fig.~\ref{fig:amps}(a).
The maximal fluctuation allowed in a AMPS depends on $N_\mathrm{aux}$, the number of the auxiliary indices.
For example, for spin-$1/2$ systems, the maximal fluctuation of total $S_z$ is $N_\mathrm{aux}$, where each auxiliary index contributes fluctuation of $1$, from $-1/2$ to $+1/2$.
The larger the $N_\mathrm{aux}$, the bigger QN jump can be achieved in each step.
However for any $N_\mathrm{aux}\geq 2$, all the QN sectors can be visited with sufficient sampling, and will all converge to the same grand canonical results.
Fig.~\ref{fig:sz} shows the convergence of total $\langle S_z\rangle$ for a $L=64$ Heisenberg chain ($H=\sum_i \mathbf{S}_i \cdot\mathbf{S}_{i+1}$) at the inverse temperature $\beta=2$, by using $N_\mathrm{aux}=2$ and a fully polarized initial state.
The convergence of total $S_z$ to zero demonstrates the ability of changing QN and simulating the grand canonical ensemble.

The use of QN not only reduces the computation cost, but also improves the Monte Carlo statistics.
The identities in the configurations represent the pre-sum in the decomposition of the partition function, and thus each configuration in the new ensemble is more efficient than the original one.
For example if $N_\mathrm{aux}=N-1$, which means only one site is collapsed, the configurations is determined by two spin degrees of freedom (one at $\tau=0$ and the other one at $\tau=\beta/2$).
The degrees of freedom of the other spins are effectively pre-summed by the identities.
The pre-sum also reduces the autocorrelation time and thus improve the convergence.

On the other hand, introducing of auxiliary indices will increase the computational cost in two ways.
The first happens in the singular value decomposition (SVD) of a two-site tensor when it consists of two auxiliary indices.
In such a case the complexity of the SVD will be $m^3d^6$ rather than $m^3d^3$, where $m$ is the bond dimension and $d$ is the physical dimension.
Another way is that the extra degree of freedom introduced from the auxiliary indices enhances the entanglement and thus increase the bond dimension of the AMPS.
The larger the $N_\mathrm{aux}$, the larger bond dimension the AMPS will have.
We will show in the result section how the computational saving outweigh the cost.

We argue that the present method is more efficient than the one in Ref~\cite{PhysRevB.95.195148}.
In Ref~\cite{PhysRevB.95.195148}, the MPS of different QN sectors are operated (time evolved, measured) separately, while in the present method they are encoded in a single AMPS and thus can likely be compressed.
Furthermore, the present method has the pre-sum representation, and thus the sampling will be more efficient.

\subsection{Connection to purification}
Purification is another finite-temperature method based on MPS~\cite{PhysRevLett.93.207204,PhysRevLett.93.207205,PhysRevB.72.220401}.
Although the purification is often represented as a combination of physical system and bath, it can be also understood as a density matrix operator, where the bath sites are understood as the "bra" (or auxiliary) sites.
The common choice of the initial purification, a product of singlet states, can be unitary transformed to a product of identities.
Therefore the algorithm can be seen as imaginary-time evolving the infinite-temperature density matrix (the product of identities).

In our new method, if we choose $N_\mathrm{aux}=N$, the initial AMPS will be the product of identities, and the algorithm will become the purification algorithm (in one Monte Carlo step).
Our method thus provides a nice connection between the METTS and the purification methods.

\begin{figure*}
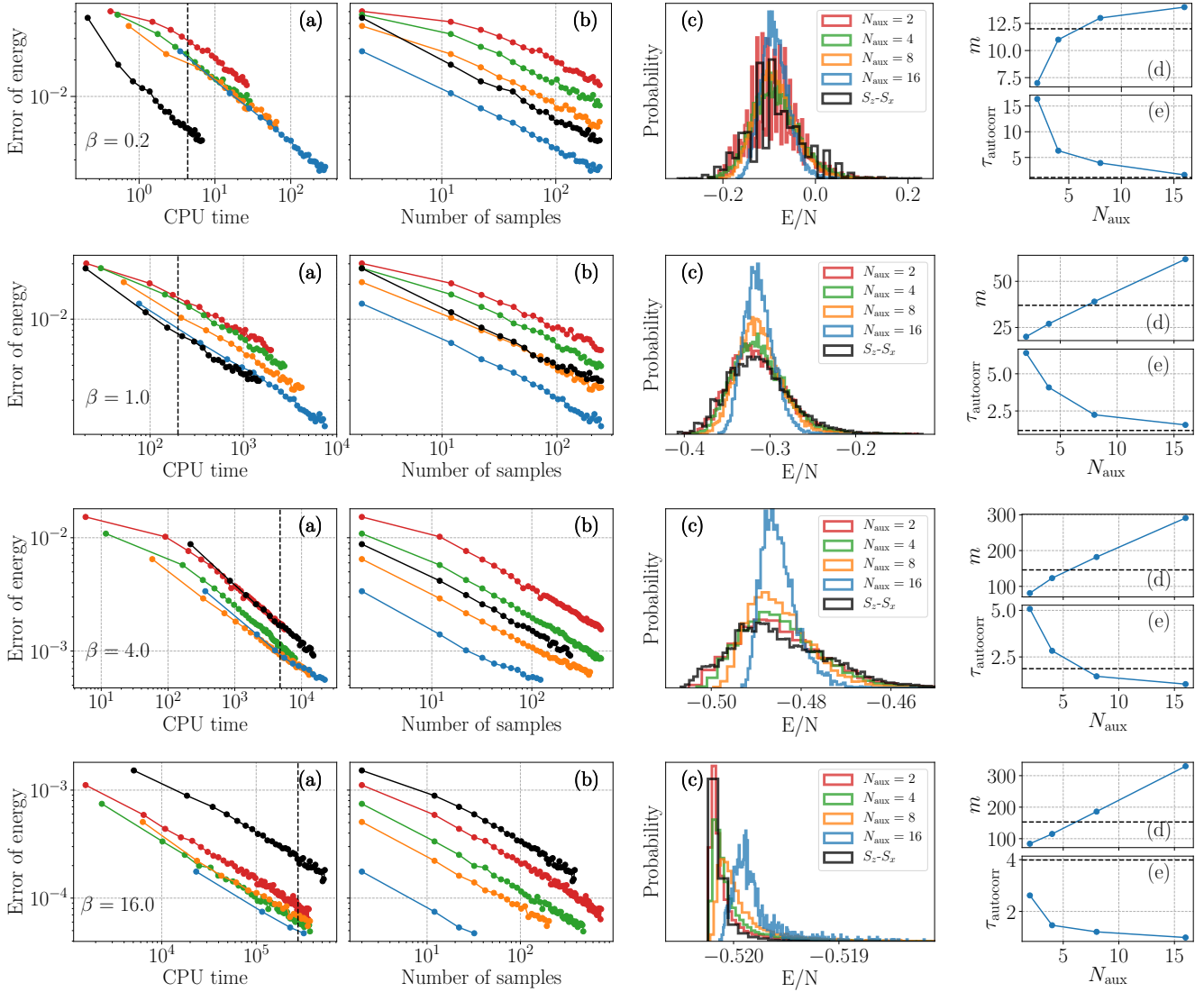

\includegraphics[scale=0.365]{{{L12x3_beta02_err_sample}}}
\includegraphics[scale=0.365]{{{L12x3_beta02_hist}}}
\includegraphics[scale=0.365]{{{L12x3_beta02_m}}}
\includegraphics[scale=0.365]{{{L12x3_beta1_err_sample}}}
\includegraphics[scale=0.365]{{{L12x3_beta1_hist}}}
\includegraphics[scale=0.365]{{{L12x3_beta1_m}}}
\includegraphics[scale=0.365]{{{L12x3_beta4_err_sample}}}
\includegraphics[scale=0.365]{{{L12x3_beta4_hist}}}
\includegraphics[scale=0.365]{{{L12x3_beta4_m}}}
\includegraphics[scale=0.365]{{{L12x3_beta16_err_sample}}}
\includegraphics[scale=0.365]{{{L12x3_beta16_hist}}}
\includegraphics[scale=0.365]{{{L12x3_beta16_m}}}
\caption
{
    Different rows of panels are for different $\beta$, as indicated in the texts in the first column of panels.
    (a) The errors of energy per site as a function of CPU time.
    The dashed line indicates the CPU time for the purification.
    The black curves are for the $S_z$-$S_x$ bases and the color curves are for AMPS bases with different $N_\mathrm{aux}$, as indicated in the third column of panels.
    (b) The errors of energy per site as functions of the number of samplings.
    (c) Probability distributions of energy per site.
    (d) Maximum bond dimension in all the QN blocks in a MPS or AMPS.
    The dashed line is for the $S_z$-$S_x$ bases. In such a case the MPS does not conserve QN and thus the maximum bond dimension will be the full bond dimension.
    (e) Autocorrelation times. The dash line is for the $S_z$-$S_x$ bases.
}
\label{fig:en}
\end{figure*}

\begin{figure*}
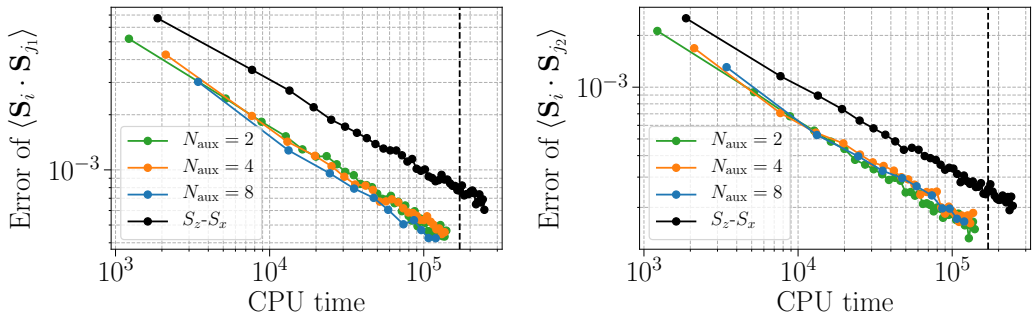

\includegraphics[width=0.8\columnwidth]{{{L12x3_beta16_corr1}}}
\includegraphics[width=0.8\columnwidth]{{{L12x3_beta16_corr2}}}
\caption
{
    The errors of the correlations $\langle S_i\cdot S_j\rangle$ for $i=(6,3)$,
    (a) $j_1=(7,3)$ and (b) $j_2=(10,1)$, where $(x,y)$ are the real-space coordinates.
}
\label{fig:corr}
\end{figure*}

\section{Benchmark results}
To demonstrate the power of our method,
we benchmark on the spin-$1/2$ XXZ model
\begin{equation}
    H = J \sum_{\langle ij\rangle} (S_i^x S_j^x + S_i^y S_j^y)
      + J_z \sum_{\langle ij\rangle} S_i^z S_j^z,
\end{equation}
on the triangular lattice.
The typical Monte Carlo methods will encounter severe sign problem due to the frustration from the lattice.
In METTS method there is clearly no sign problem because the weights are defined as positive numbers.

We choose the more difficult region $J_z/J=0.8$, where the ground state is a gapless antiferromagnetic coplanar state~\cite{doi:10.1143/JPSJ.55.3605,PhysRevLett.112.127203,PhysRevB.91.081104}.
In the original METTS algorithm, although the $S_z$-$S_x$ bases in principle requires QN-unconserved MPS, an exception exists at Heisenberg point $J_z/J=1$.
At the Heisenberg point, one can rotate the basis from $S_x$ to $S_z$ without changing the form of the Hamiltonian because of the $SU(2)$ symmetry.
In this case one needs to deal with the time evolution only for QN-conserved $S_z$-basis states~\cite{2017Benediktarxiv:1705.05578}.
However for $J_z/J\neq 1$, such a trick is no longer possible.

We consider system size $L_x\times L_y=12\times 3$ and several inverse temperatures $\beta$.
We compare the convergence of observables by using METTS with $S_z$-$S_x$ bases, METTS with AMPS bases of different $N_\mathrm{aux}$, and by using purification.
The truncation errors are controlled to $10^{-7}$ for $\beta=16$ and to $10^{-8}$ for all other $\beta$.
In the TEBD we employ a second order Suzuki-Trotter decomposition with $\delta\tau=0.1$.

\paragraph*{Energy per site.}
We first check the convergence of the energies per site, which is summarized in Fig.~\ref{fig:en}.
Different rows of panels are the results of different $\beta$, as indicated in the texts in the first column of panels.
The first (second) column of panels (Fig.~\ref{fig:en}(a (b))) shows the errors of energies as a function of CPU times (number of samplings).
The exact values of energies per site are $-0.08732$, $-0.31284$, $-0.48430$, and $-0.51978$ for $\beta=0.2, 1, 4$ and $16$ respectively.
This is the main comparison we want to make to really show the efficiencies of different methods.
In comparison of different $N_\mathrm{aux}$, although the increasing of $N_\mathrm{aux}$ will lower the errors as a function of samplings (Fig.~\ref{fig:en}(b)), they perform actually similarly when considering the CPU time (Fig.~\ref{fig:en}(a)).
This is because increasing of $N_\mathrm{aux}$ will also increase the bond dimension and thus the computational cost, so the effects balance out.
Since the small $N_\mathrm{aux}$ will generate samplings more quickly, it would be reasonable to choose small $N_\mathrm{aux}=2$ or $4$ in the general applications.

Now we compare the AMPS bases (color curves) to the $S_z$-$S_x$ bases (black curves) in Fig.~\ref{fig:en}(a).
It can be seen that, for high temperature $\beta=0.2$, the $S_z$-$S_x$ bases perform better than the AMPS bases.
At the intermediate temperature $\beta=1$ and $4$, the AMPS bases performances become compatible with the $S_z$-$S_x$ bases.
For low temperature $\beta=16$, the AMPS bases perform clearly superior to the $S_z$-$S_x$ bases.
This is significant as METTS is most useful at low temperature.

We also compare the efficiency to that of purification.
The purification simulation is done by a single imaginary time evolution and no sampling is needed.
The dashed lines in Fig.~\ref{fig:en}(a) indicate the CPU times needed in the purification simulations.
Thus the intersections to the Monte Carlo curves shows what accuracies one can obtain before the purification simulations are done.
As expected, for high temperature, the purification works most efficiently and it is not worth using METTS.
However for low temperature $\beta=16$, one can obtain four digits of accuracy, which is typically more than enough in most of the applications.
In such cases METTS is a better choice.

As mentioned in the previous Section, the new method also improves the Monte Carlo statistics.
Fig.~\ref{fig:en}(c) shows the probability histograms of the energy.
It can be seen that the introduction of the auxiliary indices narrows down the probability distribution.
The larger the $N_\mathrm{aux}$, the narrower the distribution.
This reflects the pre-sum feature in the method.
For larger $N_\mathrm{aux}$, each sampling is more important and thus more efficient.
In Fig.~\ref{fig:en}(d,e) we show the \emph{maximum} bond dimensions of \emph{every} QN blocks, and the autocorrelation time, as a function of $N_\mathrm{aux}$.
It can be seen that, when $N_\mathrm{aux}$ increases, the bond dimension increases and thus raises the computational cost, while the autocorrelation decreases and thus improves the convergence.
The bond dimensions and the autocorrelation times of the $S_z$-$S_x$ bases are shown by the dashed lines for reference.
It is interesting to point out that, for all the $\beta$, the bond dimensions of $N_\mathrm{aux}\geq 8$ have been already larger than which in the $S_z$-$S_x$ bases.
However their efficiencies are still better than the $S_z$-$S_x$ bases.
This shows that the efficiency of large $N_\mathrm{aux}$ is mainly from the efficient statistics rather than the computational gain.

\paragraph*{Correlation.}
We also show the comparisons for the correlations.
We measure the correlations $\langle\mathbf{S_i}\cdot\mathbf{S_j}\rangle$ from the center site $i=(6,3)$ to a short-distance site $j_1=(7,3)$ and a longer-distance site $j_2=(10,1)$, where $(x,y)$ are the real-space coordinates.
The exact values of the correlations are $-0.15868$ and $-0.02645$ respectively.
Fig.~\ref{fig:corr} shows the errors of the correlations as functions of CPU time, for $\beta=16$.
Again the new method of all $N_\mathrm{aux}$ are more efficient than the $S_z$-$S_x$ bases simulations, and different $N_\mathrm{aux}$ have similar efficiencies.

\section{Conclusion}
In this work we introduce a configuration representation of the METTS method, and extend the method by introducing the AMPS bases.
This basis not only encodes the pre-summation process, but also allows us to simulate the grand canonical ensembles using the QN-conserved AMPS.
We benchmark our method on the XXZ model on the triangular lattice, and study the convergence properties of the energy and the correlation.
We show that the method outperforms the original METTS and the purification method at sufficiently low temperature which is the relevant region of applications for METTS. 
In addition, the efficiency of the simulations does not significantly depend on the number of the auxiliary indices.

We mention that the configuration representation introduced in describing the algorithm can lead to further extensions.
For example one can take more slices in the imaginary time and approaches the usual quantum Monte Carlo methods.
Although the sign problem will come back, the flexibility on the choice of the bases may give us opportunities to reduce the sign problem.

\section{Acknowledgement}
We acknowledge
support by the Deutsche Forschungsgemeinschaft (DFG, German Research Foundation) 
under Germany's Excellence Strategy 426 EXC-2111 390814868.
The implementation of the algorithm uses the ITensor C++ library (version
2.1.1), https://itensor.org/.

\bibliography{qcmetts}

\begin{thebibliography}{32}
\expandafter\ifx\csname natexlab\endcsname\relax\def\natexlab#1{#1}\fi
\expandafter\ifx\csname bibnamefont\endcsname\relax
  \def\bibnamefont#1{#1}\fi
\expandafter\ifx\csname bibfnamefont\endcsname\relax
  \def\bibfnamefont#1{#1}\fi
\expandafter\ifx\csname citenamefont\endcsname\relax
  \def\citenamefont#1{#1}\fi
\expandafter\ifx\csname url\endcsname\relax
  \def\url#1{\texttt{#1}}\fi
\expandafter\ifx\csname urlprefix\endcsname\relax\def\urlprefix{URL }\fi
\providecommand{\bibinfo}[2]{#2}
\providecommand{\eprint}[2][]{\url{#2}}

\bibitem[{\citenamefont{White}(1992)}]{PhysRevLett.69.2863}
\bibinfo{author}{\bibfnamefont{S.~R.} \bibnamefont{White}},
  \bibinfo{journal}{Phys. Rev. Lett.} \textbf{\bibinfo{volume}{69}},
  \bibinfo{pages}{2863} (\bibinfo{year}{1992}),
  \urlprefix\url{https://link.aps.org/doi/10.1103/PhysRevLett.69.2863}.

\bibitem[{\citenamefont{White}(1993)}]{PhysRevB.48.10345}
\bibinfo{author}{\bibfnamefont{S.~R.} \bibnamefont{White}},
  \bibinfo{journal}{Phys. Rev. B} \textbf{\bibinfo{volume}{48}},
  \bibinfo{pages}{10345} (\bibinfo{year}{1993}),
  \urlprefix\url{https://link.aps.org/doi/10.1103/PhysRevB.48.10345}.

\bibitem[{\citenamefont{Schollw\"ock}(2005)}]{RevModPhys.77.259}
\bibinfo{author}{\bibfnamefont{U.}~\bibnamefont{Schollw\"ock}},
  \bibinfo{journal}{Rev. Mod. Phys.} \textbf{\bibinfo{volume}{77}},
  \bibinfo{pages}{259} (\bibinfo{year}{2005}),
  \urlprefix\url{https://link.aps.org/doi/10.1103/RevModPhys.77.259}.

\bibitem[{\citenamefont{McCulloch}(2007)}]{McCulloch_2007}
\bibinfo{author}{\bibfnamefont{I.~P.} \bibnamefont{McCulloch}},
  \bibinfo{journal}{Journal of Statistical Mechanics: Theory and Experiment}
  \textbf{\bibinfo{volume}{2007}}, \bibinfo{pages}{P10014}
  (\bibinfo{year}{2007}),
  \urlprefix\url{https://doi.org/10.1088%2F1742-5468%2F2007%2F10%2Fp10014}.

\bibitem[{\citenamefont{Schollw\"ock}(2011)}]{Schollw_ck_2011}
\bibinfo{author}{\bibfnamefont{U.}~\bibnamefont{Schollw\"ock}},
  \bibinfo{journal}{Annals of Physics} \textbf{\bibinfo{volume}{326}},
  \bibinfo{pages}{96} (\bibinfo{year}{2011}),
  \urlprefix\url{https://doi.org/10.1016%2Fj.aop.2010.09.012}.

\bibitem[{\citenamefont{White}(2009)}]{PhysRevLett.102.190601}
\bibinfo{author}{\bibfnamefont{S.~R.} \bibnamefont{White}},
  \bibinfo{journal}{Phys. Rev. Lett.} \textbf{\bibinfo{volume}{102}},
  \bibinfo{pages}{190601} (\bibinfo{year}{2009}),
  \urlprefix\url{https://link.aps.org/doi/10.1103/PhysRevLett.102.190601}.

\bibitem[{\citenamefont{Stoudenmire and White}(2010)}]{Stoudenmire_2010}
\bibinfo{author}{\bibfnamefont{E.~M.} \bibnamefont{Stoudenmire}}
  \bibnamefont{and} \bibinfo{author}{\bibfnamefont{S.~R.} \bibnamefont{White}},
  \bibinfo{journal}{New Journal of Physics} \textbf{\bibinfo{volume}{12}},
  \bibinfo{pages}{055026} (\bibinfo{year}{2010}),
  \urlprefix\url{https://doi.org/10.1088%2F1367-2630%2F12%2F5%2F055026}.

\bibitem[{\citenamefont{Bruognolo et~al.}(2015)\citenamefont{Bruognolo, von
  Delft, and Weichselbaum}}]{PhysRevB.92.115105}
\bibinfo{author}{\bibfnamefont{B.}~\bibnamefont{Bruognolo}},
  \bibinfo{author}{\bibfnamefont{J.}~\bibnamefont{von Delft}},
  \bibnamefont{and}
  \bibinfo{author}{\bibfnamefont{A.}~\bibnamefont{Weichselbaum}},
  \bibinfo{journal}{Phys. Rev. B} \textbf{\bibinfo{volume}{92}},
  \bibinfo{pages}{115105} (\bibinfo{year}{2015}),
  \urlprefix\url{https://link.aps.org/doi/10.1103/PhysRevB.92.115105}.

\bibitem[{\citenamefont{Verstraete et~al.}(2004)\citenamefont{Verstraete,
  Garc\'{\i}a-Ripoll, and Cirac}}]{PhysRevLett.93.207204}
\bibinfo{author}{\bibfnamefont{F.}~\bibnamefont{Verstraete}},
  \bibinfo{author}{\bibfnamefont{J.~J.} \bibnamefont{Garc\'{\i}a-Ripoll}},
  \bibnamefont{and} \bibinfo{author}{\bibfnamefont{J.~I.} \bibnamefont{Cirac}},
  \bibinfo{journal}{Phys. Rev. Lett.} \textbf{\bibinfo{volume}{93}},
  \bibinfo{pages}{207204} (\bibinfo{year}{2004}),
  \urlprefix\url{https://link.aps.org/doi/10.1103/PhysRevLett.93.207204}.

\bibitem[{\citenamefont{Zwolak and Vidal}(2004)}]{PhysRevLett.93.207205}
\bibinfo{author}{\bibfnamefont{M.}~\bibnamefont{Zwolak}} \bibnamefont{and}
  \bibinfo{author}{\bibfnamefont{G.}~\bibnamefont{Vidal}},
  \bibinfo{journal}{Phys. Rev. Lett.} \textbf{\bibinfo{volume}{93}},
  \bibinfo{pages}{207205} (\bibinfo{year}{2004}),
  \urlprefix\url{https://link.aps.org/doi/10.1103/PhysRevLett.93.207205}.

\bibitem[{\citenamefont{Feiguin and White}(2005)}]{PhysRevB.72.220401}
\bibinfo{author}{\bibfnamefont{A.~E.} \bibnamefont{Feiguin}} \bibnamefont{and}
  \bibinfo{author}{\bibfnamefont{S.~R.} \bibnamefont{White}},
  \bibinfo{journal}{Phys. Rev. B} \textbf{\bibinfo{volume}{72}},
  \bibinfo{pages}{220401} (\bibinfo{year}{2005}),
  \urlprefix\url{https://link.aps.org/doi/10.1103/PhysRevB.72.220401}.

\bibitem[{\citenamefont{Nocera and Alvarez}(2016)}]{PhysRevB.93.045137}
\bibinfo{author}{\bibfnamefont{A.}~\bibnamefont{Nocera}} \bibnamefont{and}
  \bibinfo{author}{\bibfnamefont{G.}~\bibnamefont{Alvarez}},
  \bibinfo{journal}{Phys. Rev. B} \textbf{\bibinfo{volume}{93}},
  \bibinfo{pages}{045137} (\bibinfo{year}{2016}),
  \urlprefix\url{https://link.aps.org/doi/10.1103/PhysRevB.93.045137}.

\bibitem[{\citenamefont{Barthel}(2016)}]{PhysRevB.94.115157}
\bibinfo{author}{\bibfnamefont{T.}~\bibnamefont{Barthel}},
  \bibinfo{journal}{Phys. Rev. B} \textbf{\bibinfo{volume}{94}},
  \bibinfo{pages}{115157} (\bibinfo{year}{2016}),
  \urlprefix\url{https://link.aps.org/doi/10.1103/PhysRevB.94.115157}.

\bibitem[{\citenamefont{Hauschild et~al.}(2018)\citenamefont{Hauschild,
  Leviatan, Bardarson, Altman, Zaletel, and Pollmann}}]{PhysRevB.98.235163}
\bibinfo{author}{\bibfnamefont{J.}~\bibnamefont{Hauschild}},
  \bibinfo{author}{\bibfnamefont{E.}~\bibnamefont{Leviatan}},
  \bibinfo{author}{\bibfnamefont{J.~H.} \bibnamefont{Bardarson}},
  \bibinfo{author}{\bibfnamefont{E.}~\bibnamefont{Altman}},
  \bibinfo{author}{\bibfnamefont{M.~P.} \bibnamefont{Zaletel}},
  \bibnamefont{and} \bibinfo{author}{\bibfnamefont{F.}~\bibnamefont{Pollmann}},
  \bibinfo{journal}{Phys. Rev. B} \textbf{\bibinfo{volume}{98}},
  \bibinfo{pages}{235163} (\bibinfo{year}{2018}),
  \urlprefix\url{https://link.aps.org/doi/10.1103/PhysRevB.98.235163}.

\bibitem[{\citenamefont{Chen et~al.}(2018)\citenamefont{Chen, Chen, Chen, Li,
  and Weichselbaum}}]{PhysRevX.8.031082}
\bibinfo{author}{\bibfnamefont{B.-B.} \bibnamefont{Chen}},
  \bibinfo{author}{\bibfnamefont{L.}~\bibnamefont{Chen}},
  \bibinfo{author}{\bibfnamefont{Z.}~\bibnamefont{Chen}},
  \bibinfo{author}{\bibfnamefont{W.}~\bibnamefont{Li}}, \bibnamefont{and}
  \bibinfo{author}{\bibfnamefont{A.}~\bibnamefont{Weichselbaum}},
  \bibinfo{journal}{Phys. Rev. X} \textbf{\bibinfo{volume}{8}},
  \bibinfo{pages}{031082} (\bibinfo{year}{2018}),
  \urlprefix\url{https://link.aps.org/doi/10.1103/PhysRevX.8.031082}.

\bibitem[{\citenamefont{Li et~al.}(2019)\citenamefont{Li, Chen, Chen, von
  Delft, Weichselbaum, and Li}}]{PhysRevB.100.045110}
\bibinfo{author}{\bibfnamefont{H.}~\bibnamefont{Li}},
  \bibinfo{author}{\bibfnamefont{B.-B.} \bibnamefont{Chen}},
  \bibinfo{author}{\bibfnamefont{Z.}~\bibnamefont{Chen}},
  \bibinfo{author}{\bibfnamefont{J.}~\bibnamefont{von Delft}},
  \bibinfo{author}{\bibfnamefont{A.}~\bibnamefont{Weichselbaum}},
  \bibnamefont{and} \bibinfo{author}{\bibfnamefont{W.}~\bibnamefont{Li}},
  \bibinfo{journal}{Phys. Rev. B} \textbf{\bibinfo{volume}{100}},
  \bibinfo{pages}{045110} (\bibinfo{year}{2019}),
  \urlprefix\url{https://link.aps.org/doi/10.1103/PhysRevB.100.045110}.

\bibitem[{\citenamefont{Shibata}(1997)}]{doi:10.1143/JPSJ.66.2221}
\bibinfo{author}{\bibfnamefont{N.}~\bibnamefont{Shibata}},
  \bibinfo{journal}{Journal of the Physical Society of Japan}
  \textbf{\bibinfo{volume}{66}}, \bibinfo{pages}{2221} (\bibinfo{year}{1997}),
  \eprint{https://doi.org/10.1143/JPSJ.66.2221},
  \urlprefix\url{https://doi.org/10.1143/JPSJ.66.2221}.

\bibitem[{\citenamefont{Nishino}(1995)}]{doi:10.1143/JPSJ.64.3598}
\bibinfo{author}{\bibfnamefont{T.}~\bibnamefont{Nishino}},
  \bibinfo{journal}{Journal of the Physical Society of Japan}
  \textbf{\bibinfo{volume}{64}}, \bibinfo{pages}{3598} (\bibinfo{year}{1995}),
  \eprint{https://doi.org/10.1143/JPSJ.64.3598},
  \urlprefix\url{https://doi.org/10.1143/JPSJ.64.3598}.

\bibitem[{\citenamefont{Bursill et~al.}(1996)\citenamefont{Bursill, Xiang, and
  Gehring}}]{Bursill_1996}
\bibinfo{author}{\bibfnamefont{R.~J.} \bibnamefont{Bursill}},
  \bibinfo{author}{\bibfnamefont{T.}~\bibnamefont{Xiang}}, \bibnamefont{and}
  \bibinfo{author}{\bibfnamefont{G.~A.} \bibnamefont{Gehring}},
  \bibinfo{journal}{Journal of Physics: Condensed Matter}
  \textbf{\bibinfo{volume}{8}}, \bibinfo{pages}{L583} (\bibinfo{year}{1996}),
  \urlprefix\url{https://doi.org/10.1088%2F0953-8984%2F8%2F40%2F003}.

\bibitem[{\citenamefont{Wang and Xiang}(1997)}]{PhysRevB.56.5061}
\bibinfo{author}{\bibfnamefont{X.}~\bibnamefont{Wang}} \bibnamefont{and}
  \bibinfo{author}{\bibfnamefont{T.}~\bibnamefont{Xiang}},
  \bibinfo{journal}{Phys. Rev. B} \textbf{\bibinfo{volume}{56}},
  \bibinfo{pages}{5061} (\bibinfo{year}{1997}),
  \urlprefix\url{https://link.aps.org/doi/10.1103/PhysRevB.56.5061}.

\bibitem[{\citenamefont{Binder and Barthel}(2015)}]{PhysRevB.92.125119}
\bibinfo{author}{\bibfnamefont{M.}~\bibnamefont{Binder}} \bibnamefont{and}
  \bibinfo{author}{\bibfnamefont{T.}~\bibnamefont{Barthel}},
  \bibinfo{journal}{Phys. Rev. B} \textbf{\bibinfo{volume}{92}},
  \bibinfo{pages}{125119} (\bibinfo{year}{2015}),
  \urlprefix\url{https://link.aps.org/doi/10.1103/PhysRevB.92.125119}.

\bibitem[{\citenamefont{Binder and Barthel}(2017)}]{PhysRevB.95.195148}
\bibinfo{author}{\bibfnamefont{M.}~\bibnamefont{Binder}} \bibnamefont{and}
  \bibinfo{author}{\bibfnamefont{T.}~\bibnamefont{Barthel}},
  \bibinfo{journal}{Phys. Rev. B} \textbf{\bibinfo{volume}{95}},
  \bibinfo{pages}{195148} (\bibinfo{year}{2017}),
  \urlprefix\url{https://link.aps.org/doi/10.1103/PhysRevB.95.195148}.

\bibitem[{\citenamefont{Vidal}(2004)}]{PhysRevLett.93.040502}
\bibinfo{author}{\bibfnamefont{G.}~\bibnamefont{Vidal}},
  \bibinfo{journal}{Phys. Rev. Lett.} \textbf{\bibinfo{volume}{93}},
  \bibinfo{pages}{040502} (\bibinfo{year}{2004}),
  \urlprefix\url{https://link.aps.org/doi/10.1103/PhysRevLett.93.040502}.

\bibitem[{\citenamefont{White and Feiguin}(2004)}]{PhysRevLett.93.076401}
\bibinfo{author}{\bibfnamefont{S.~R.} \bibnamefont{White}} \bibnamefont{and}
  \bibinfo{author}{\bibfnamefont{A.~E.} \bibnamefont{Feiguin}},
  \bibinfo{journal}{Phys. Rev. Lett.} \textbf{\bibinfo{volume}{93}},
  \bibinfo{pages}{076401} (\bibinfo{year}{2004}),
  \urlprefix\url{https://link.aps.org/doi/10.1103/PhysRevLett.93.076401}.

\bibitem[{\citenamefont{Daley et~al.}(2004)\citenamefont{Daley, Kollath,
  Schollwöck, and Vidal}}]{Daley_2004}
\bibinfo{author}{\bibfnamefont{A.~J.} \bibnamefont{Daley}},
  \bibinfo{author}{\bibfnamefont{C.}~\bibnamefont{Kollath}},
  \bibinfo{author}{\bibfnamefont{U.}~\bibnamefont{Schollwöck}},
  \bibnamefont{and} \bibinfo{author}{\bibfnamefont{G.}~\bibnamefont{Vidal}},
  \bibinfo{journal}{Journal of Statistical Mechanics: Theory and Experiment}
  \textbf{\bibinfo{volume}{2004}}, \bibinfo{pages}{P04005}
  (\bibinfo{year}{2004}),
  \urlprefix\url{https://doi.org/10.1088%2F1742-5468%2F2004%2F04%2Fp04005}.

\bibitem[{\citenamefont{Haegeman et~al.}(2011)\citenamefont{Haegeman, Cirac,
  Osborne, Pi\ifmmode~\check{z}\else \v{z}\fi{}orn, Verschelde, and
  Verstraete}}]{PhysRevLett.107.070601}
\bibinfo{author}{\bibfnamefont{J.}~\bibnamefont{Haegeman}},
  \bibinfo{author}{\bibfnamefont{J.~I.} \bibnamefont{Cirac}},
  \bibinfo{author}{\bibfnamefont{T.~J.} \bibnamefont{Osborne}},
  \bibinfo{author}{\bibfnamefont{I.}~\bibnamefont{Pi\ifmmode~\check{z}\else
  \v{z}\fi{}orn}},
  \bibinfo{author}{\bibfnamefont{H.}~\bibnamefont{Verschelde}},
  \bibnamefont{and}
  \bibinfo{author}{\bibfnamefont{F.}~\bibnamefont{Verstraete}},
  \bibinfo{journal}{Phys. Rev. Lett.} \textbf{\bibinfo{volume}{107}},
  \bibinfo{pages}{070601} (\bibinfo{year}{2011}),
  \urlprefix\url{https://link.aps.org/doi/10.1103/PhysRevLett.107.070601}.

\bibitem[{\citenamefont{Haegeman et~al.}(2016)\citenamefont{Haegeman, Lubich,
  Oseledets, Vandereycken, and Verstraete}}]{PhysRevB.94.165116}
\bibinfo{author}{\bibfnamefont{J.}~\bibnamefont{Haegeman}},
  \bibinfo{author}{\bibfnamefont{C.}~\bibnamefont{Lubich}},
  \bibinfo{author}{\bibfnamefont{I.}~\bibnamefont{Oseledets}},
  \bibinfo{author}{\bibfnamefont{B.}~\bibnamefont{Vandereycken}},
  \bibnamefont{and}
  \bibinfo{author}{\bibfnamefont{F.}~\bibnamefont{Verstraete}},
  \bibinfo{journal}{Phys. Rev. B} \textbf{\bibinfo{volume}{94}},
  \bibinfo{pages}{165116} (\bibinfo{year}{2016}),
  \urlprefix\url{https://link.aps.org/doi/10.1103/PhysRevB.94.165116}.

\bibitem[{\citenamefont{{Paeckel} et~al.}(2019)\citenamefont{{Paeckel},
  {K{\"o}hler}, {Swoboda}, {Manmana}, {Schollw{\"o}ck}, and
  {Hubig}}}]{2019arXiv190105824P}
\bibinfo{author}{\bibfnamefont{S.}~\bibnamefont{{Paeckel}}},
  \bibinfo{author}{\bibfnamefont{T.}~\bibnamefont{{K{\"o}hler}}},
  \bibinfo{author}{\bibfnamefont{A.}~\bibnamefont{{Swoboda}}},
  \bibinfo{author}{\bibfnamefont{S.~R.} \bibnamefont{{Manmana}}},
  \bibinfo{author}{\bibfnamefont{U.}~\bibnamefont{{Schollw{\"o}ck}}},
  \bibnamefont{and} \bibinfo{author}{\bibfnamefont{C.}~\bibnamefont{{Hubig}}},
  \bibinfo{journal}{arXiv e-prints} \bibinfo{eid}{arXiv:1901.05824}
  (\bibinfo{year}{2019}), \eprint{1901.05824}.

\bibitem[{\citenamefont{Miyashita}(1986)}]{doi:10.1143/JPSJ.55.3605}
\bibinfo{author}{\bibfnamefont{S.}~\bibnamefont{Miyashita}},
  \bibinfo{journal}{Journal of the Physical Society of Japan}
  \textbf{\bibinfo{volume}{55}}, \bibinfo{pages}{3605} (\bibinfo{year}{1986}),
  \eprint{https://doi.org/10.1143/JPSJ.55.3605},
  \urlprefix\url{https://doi.org/10.1143/JPSJ.55.3605}.

\bibitem[{\citenamefont{Yamamoto et~al.}(2014)\citenamefont{Yamamoto,
  Marmorini, and Danshita}}]{PhysRevLett.112.127203}
\bibinfo{author}{\bibfnamefont{D.}~\bibnamefont{Yamamoto}},
  \bibinfo{author}{\bibfnamefont{G.}~\bibnamefont{Marmorini}},
  \bibnamefont{and} \bibinfo{author}{\bibfnamefont{I.}~\bibnamefont{Danshita}},
  \bibinfo{journal}{Phys. Rev. Lett.} \textbf{\bibinfo{volume}{112}},
  \bibinfo{pages}{127203} (\bibinfo{year}{2014}),
  \urlprefix\url{https://link.aps.org/doi/10.1103/PhysRevLett.112.127203}.

\bibitem[{\citenamefont{Sellmann et~al.}(2015)\citenamefont{Sellmann, Zhang,
  and Eggert}}]{PhysRevB.91.081104}
\bibinfo{author}{\bibfnamefont{D.}~\bibnamefont{Sellmann}},
  \bibinfo{author}{\bibfnamefont{X.-F.} \bibnamefont{Zhang}}, \bibnamefont{and}
  \bibinfo{author}{\bibfnamefont{S.}~\bibnamefont{Eggert}},
  \bibinfo{journal}{Phys. Rev. B} \textbf{\bibinfo{volume}{91}},
  \bibinfo{pages}{081104} (\bibinfo{year}{2015}),
  \urlprefix\url{https://link.aps.org/doi/10.1103/PhysRevB.91.081104}.

\bibitem[{\citenamefont{Bruognolo et~al.}(2017)\citenamefont{Bruognolo, Zhu,
  White, and Stoudenmire}}]{2017Benediktarxiv:1705.05578}
\bibinfo{author}{\bibfnamefont{B.}~\bibnamefont{Bruognolo}},
  \bibinfo{author}{\bibfnamefont{Z.}~\bibnamefont{Zhu}},
  \bibinfo{author}{\bibfnamefont{S.~R.} \bibnamefont{White}}, \bibnamefont{and}
  \bibinfo{author}{\bibfnamefont{E.~M.} \bibnamefont{Stoudenmire}},
  \bibinfo{journal}{arxiv:1705.05578}  (\bibinfo{year}{2017}),
  \urlprefix\url{http://arxiv.org/abs/1705.05578v2}.

\end{thebibliography}

\end{document}